\documentclass[a4paper,11pt]{article}
\usepackage{pos}

\title{Modelling of highly extended Gamma-ray emission around the Geminga Pulsar as detected with H.E.S.S.}
 \ShortTitle{Modelling Extended Gamma-ray emission around Geminga}

\author*[a]{A. M. W. Mitchell}
\author[b]{S. Caroff}

\affiliation[a]{Erlangen Centre for Astroparticle Physics, Friedrich-Alexander-Universit\"at Erlangen-N\"urnberg,\\ 
Nikolaus-Fiebiger-Str. 2, 91058 Erlangen, Germany}

\affiliation[b]{Laboratoire d'Annecy de Physique des Particules, Univ. Grenoble Alpes, \\
Univ. Savoie Mont Blanc, CNRS, LAPP, 74000 Annecy, France}

\onbehalf{for the H.E.S.S. collaboration} 

\emailAdd{contact.hess@hess-experiment.eu}
\emailAdd{alison.mw.mitchell@fau.de}
\emailAdd{sami.caroff@lapp.in2p3.fr}

\abstract{Geminga is an enigmatic radio-quiet gamma-ray pulsar located at a mere 250 pc distance from Earth. Extended very-high-energy gamma-ray emission around the pulsar has been detected by multiple water Cherenkov detector based instruments. However, the detection of extended TeV gamma-ray emission around the Geminga pulsar has proven challenging for IACTs due to the angular scale exceeding the typical field-of-view. By detailed studies of background estimation techniques and characterising systematic effects, a detection of highly extended TeV gamma-ray emission could be confirmed by the H.E.S.S. IACT array. Building on the previously announced detection, in this contribution we further characterise the emission and apply an electron diffusion model to the combined gamma-ray data from the H.E.S.S. and HAWC experiments, as well as X-ray data from XMM-Newton.}

\ConferenceLogo{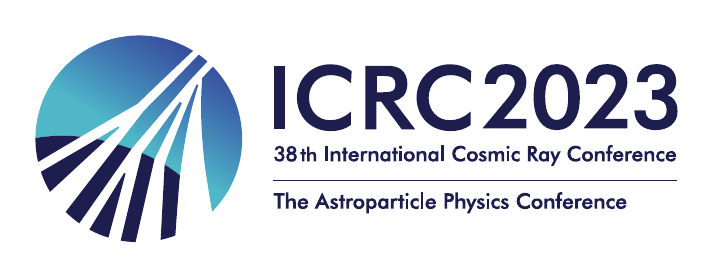}

\FullConference{%
38th International Cosmic Ray Conference (ICRC2023)\\
  26 July - 3 August, 2023\\
  Nagoya, Japan}

\begin{document}
\maketitle

\section{Introduction}

Geminga (PSR\,J0633+1746) is a middle-aged pulsar ($\tau_c=342$\,kyr) in close proximity to Earth ($d=250$\,pc), with a spin-down luminosity of $\dot{E}=3.2\times10^{34}$erg/s, that is radio quiet, yet exhibits pulsed gamma-ray emission. The detection of extended $\gamma$-ray emission coincident with the pulsar was first achieved by Milagro \cite{Milagro07}
and subsequently verified by HAWC \cite{hawc}, yet the angular scale of $\gtrsim2^\circ$ posed a challenge for Imaging Atmospheric Cherenkov Telescopes (IACTs). The angular scale of the very-high-energy (VHE, $E\gtrsim100$\,GeV) $\gamma$-ray emission of $\sim5.5^\circ$ is considerably larger than that of the associated X-ray pulsar wind nebula (PWN), of $\sim10'$ \cite{Geminga_Xray}.

Given that the majority of PWNe that are detected in VHE $\gamma$-rays are associated with young, energetic pulsars and that at these later stages the structure of the former PWN has been disrupted, such that particles can leak out into the surrounding interstellar medium (ISM), it was proposed that the $\gamma$-ray emission sounding Geminga (and the nearby companion PSR\,B0656+14) form a distinct class in the evolutionary history of pulsar environments, termed `pulsar halos' (or `TeV halos', whereby the latter is a popular yet ambiguous term) \cite{Giacinti_halo,Linden2017PhRvD..96j3016L}. 
A key distinguishing feature between PWNe and pulsar halos is the average energy density in electrons responsible for the $\gamma$-ray emission via inverse Compton scattering, which for PWNe is higher and for halos lower than that typical of the surrounding ISM \cite{Giacinti_halo,psrhaloreview}. %O(1eV/cm^3)

With improving performance and exposure of ground-based particle detector facilities such as HAWC and LHAASO, the $\gamma$-ray sky has continued to reveal an increasing number of pulsar halo systems \cite{psrhaloreview}.
The morphology of the emission detected with HAWC around the Geminga pulsar indicated that the diffusion coefficient in the vicinity of the pulsar is a factor $\sim100$ below the Galactic average expected for the ISM. Several scenarios have been suggested to reconcile the two, such as suppressed diffusion due to turbulence in the vicinity of the pulsar \cite{Evoli}. 

Accounting for analysis differences between experiments, H.E.S.S. was able to detect the presence of extended $\gamma$-ray emission around the Geminga pulsar \cite{hess,hawchess}. 
To adjust for the large angular size, an observation campaign was conducted in 2019 with telescope pointing offsets of $1.6^\circ$ (much larger than the usual $\sim0.7^\circ$), %this is a bit of iact slang - ok? 
from which a detailed analysis and modelling could be performed. These proceedings provide a summary of the key analysis results and focus on the modelling, where we endeavour to perform a joint fit combining data from HAWC and XMM-Newton to place constraints on the diffusion properties. 

\section{H.E.S.S. Data Analysis}

\begin{figure}
\centering
\includegraphics[width=0.5\textwidth]{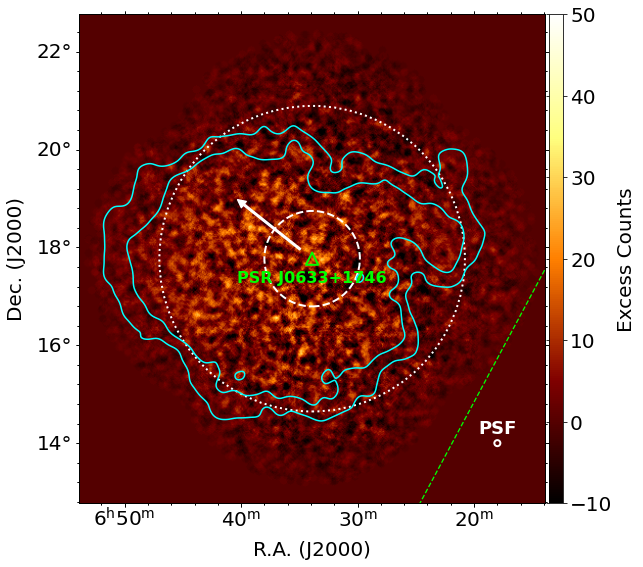}
\caption{Excess counts sky map of the region around the Geminga pulsar using 2019 data from the H.E.S.S. experiment, analysed with an On-Off background method \cite{hess}. The location of the Geminga pulsar is indicated with a green triangle. White dashed and dotted circles indicate the $1^\circ$ and $3.2^\circ$ radius regions used for the spectral analysis and the radial profile respectively.  }
\label{fig:map}
\end{figure}

In \cite{hess}, the H.E.S.S. Collaboration reported the significant detection of extended gamma-ray emission around the Geminga pulsar, out to at least $3.2^\circ$ radius. An excess counts sky map constructed using the On-Off background estimation method is shown in figure \ref{fig:map}. The 2019 dataset provided 27.2\,h exposure with observations obtained at offsets of $\pm1.6^\circ$ from the location of the Geminga pulsar. Background normalisation was hence performed on data beyond $3.2^\circ$ (twice the angular pointing offset). This limitation to the sky region meant that the full extent of the emission could not be measured, yet a relative measurement indicating a significant excess above background level was nevertheless found. 

Within the innermost $1^\circ$, a significance of $\sim9-10\,\sigma$ was obtained with different background estimation methods. A spectral analysis was performed to this region, indicated by a white dashed line in figure \ref{fig:map}. A power law spectral model was fit to the data, $\frac{dN}{dE}=\phi_0\left(\frac{E}{E_0}\right)^{-\Gamma}$ with best-fit spectral index $\Gamma=2.76\pm 0.22$ and flux normalisation at 1\,TeV of $(2.8\pm0.7)\times10^{-12}$cm$^{-2}$s$^{-1}$TeV$^{-1}$. 
The spectral results and radial profile are shown together with best-fit models below. 

The centroid of the $\gamma$-ray emission across the energy range 0.5\,TeV -- 40\,TeV was found to be located at an offset of $0.6^\circ$ from the pulsar, at R.A. $99.1^\circ \pm 0.1^\circ \pm 0.5^\circ$ and Dec. $17.7^\circ \pm 0.1^\circ \pm 0.5^\circ$, which is nevertheless compatible with the pulsar position within the systematic errors. Evaluating the 68\% containment radii in different energy bands, no evidence for statistically significant energy-dependent morphology was found.

\section{Diffusion Model}

To describe the $\gamma$-ray emission, we consider a scenario of electrons diffusing away from the pulsar within a halo region, where the diffusion coefficient has a dependence on energy as $D(E)=D_0(E/E_0)^\delta$ with $\delta \in [0.3,1]$. The pulsar is considered as a point-like continuous source of electrons, for which we take energy-dependent diffusion and energy losses into account. We solve the diffusive transport equation:
\begin{equation}
    \partial_t N(E,\vec{r},t) - D(E)\Delta N(E,\vec{r},t) + \partial_E[b(E)N(E,\vec{r},t)]  = Q(E,t)\,\delta(\vec{r}-\vec{r_s}),
\end{equation}
where the source term $Q(E,t)$ depends on the energy released by the pulsar and describes the injection of electrons into the pulsar environment:
\begin{equation}
    Q(E,t) = Q_0 (1+t/\tau_0)^{-\frac{(n+1)}{(n-1)}} (E/E_0)^{-\alpha}\exp{(-E/E_c)}~,
\end{equation}
with an initial spin-down timescale $\tau_0$ and braking index $n$.
The solution adopted for the diffusion equation is:
\begin{equation}
    N(E,r,T_*) = \int^{T_*}_0 \mathrm{d}t_0 \frac{b(E_s(E,t_0,T_*))}{b(E)} \frac{1}{(\pi \lambda^2(t_0,T_*,E))^{3/2}} 
    \times  \exp{\left(-\frac{r^2 + r_s^2(t_0)}{\lambda^2(t_0,T_*,E)}\right)}Q(E_s(E,t_0,T_*),t_0)~,
\end{equation}
where the subscript $_*$ indicates properties of the pulsar at the current time and $\lambda$ is the diffusion length. %Anything to add here? 
Table \ref{tab:Param} summarises several parameters of the model, including their fixed and/or scanned values as appropriate. 

\begin{figure}
\centering
\includegraphics[width=0.48\textwidth]{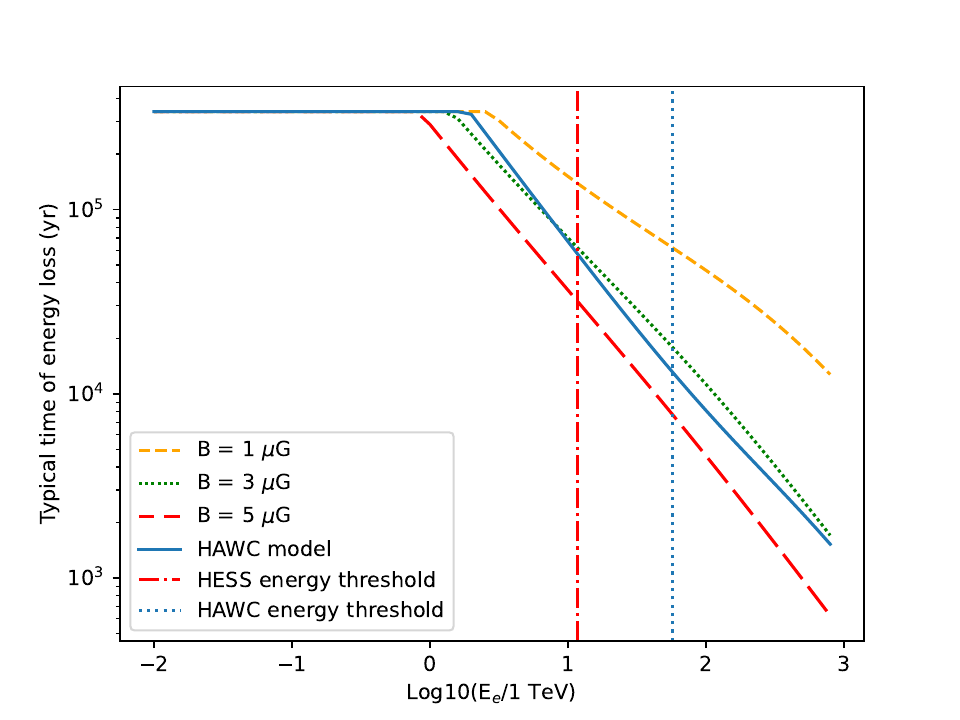}
\includegraphics[width=0.50\textwidth]{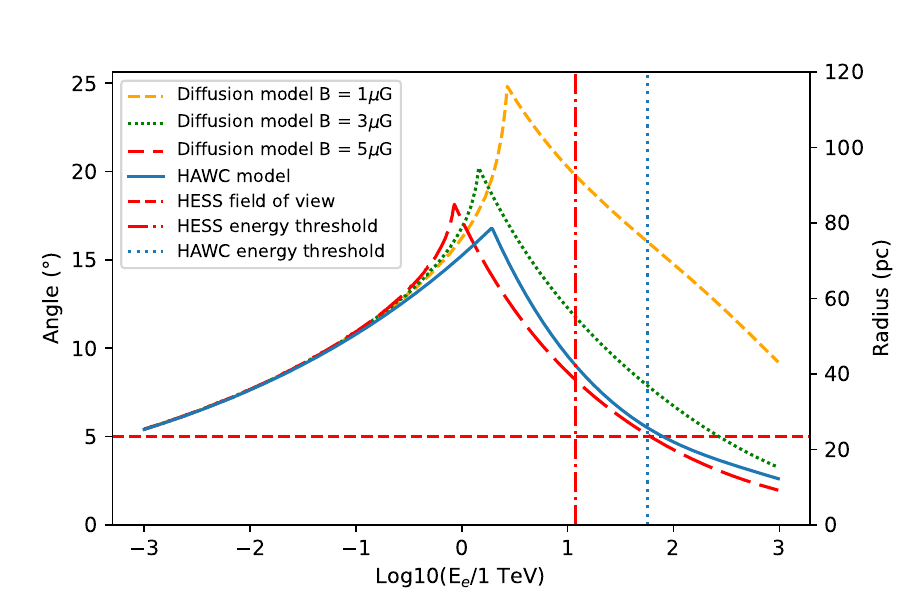}
\caption{Properties of the electron diffusion model applied. Left: electron energy loss timescale, and Right: electron diffusion radius, both with electron energy and magnetic field strength. The H.E.S.S. energy threshold of $\sim$1\,TeV applied to this analysis corresponds to an electron energy of $\gtrsim10$\,TeV. }
\label{fig:model}
\end{figure}

Figure \ref{fig:model} shows the energy loss time and the diffusion radius. Electrons with energies $\lesssim1$\,TeV have not yet cooled, as the loss timescale is larger than the age of the pulsar. Correspondingly, the peak diffusion radius also occurs at around 1\,TeV, above which the expected size due to diffusion decreases with increasing energy. At the energy threshold of H.E.S.S., the diffusion radius is larger than the field of view of the H.E.S.S. telescopes.  

\begin{table}[h!]
\caption{Input parameters for the diffusion model, where $D_0$ is the normalisation at an electron energy $E_0=100$\,TeV. Properties of the pulsar are set to known values where available (e.g. age $T_c$, distance $d$, and the luminosity and spin period at the current time, $L_*$, $P_*$)}
\vspace{-0.5cm}
\begin{center}
\begin{tabular}{ cc } 
 Parameter & Value(s) \\ 
 \hline
% Current Luminosity, $L_*$ &  $3.2 \times 10^{34}$ erg/s \\ 
%Pulsar age, $T_c$  & $342$ kyr \\ 
 Braking index, $n$  & [$1.5$, $3$, $4.5$]\\
 Initial period, $P_0$  & $15$ ms \\
 %Current period, $P_*$  & $0.237$ s \\
 %Distance, $d$  & $250$ pc\\
 Transverse velocity, $V_T$  & $211$ km/s \\
 Electron efficiency, $\eta$  & [$0.01$, $0.1$, $0.5$] \\ 
 Electron injection index, $\alpha$ & [$1.8$, $2.0$, $2.2$] \\
 Electron energy cut-off, $E_c$ & [free,$1$ PeV] \\
 Ambient magnetic field, $B$ & [$1$, $3$, $5$]  $\mu$G \\ 
 Diffusion coefficient normalisation, $D_0$  & free  \\ 
 Diffusion index, $\delta$   & [$0.3$, $0.6$, $1$]
\end{tabular}
\end{center}
\label{tab:Param}
\end{table}

\section{Modelling Results}

To obtain the best-fit model to the HAWC, H.E.S.S. and  XMM-Newton data, we performed a parameter scan over variables of the diffusion model as listed in table \ref{tab:Param}. Five variables ($n,\eta,\alpha,B,\delta$) were scanned over three values, yielding a total of 243 different parameter combinations. The normalisation of the diffusion coefficient was always left as a free parameter of the fit. A global minimisation procedure was found not to converge, as multiple parameter combinations could yield comparably consistent matches to the data. A combination of model parameters was considered a good fit to the data if a p-value of $>0.003$ was obtained, a criterion achieved by 53 out of the 243 parameter combinations. 
The process was repeated with both the cut-off energy of the electron spectrum $E_c$ fixed to 1\,PeV and left free to vary. 

\begin{figure}
\centering
\includegraphics[width=0.49\textwidth]{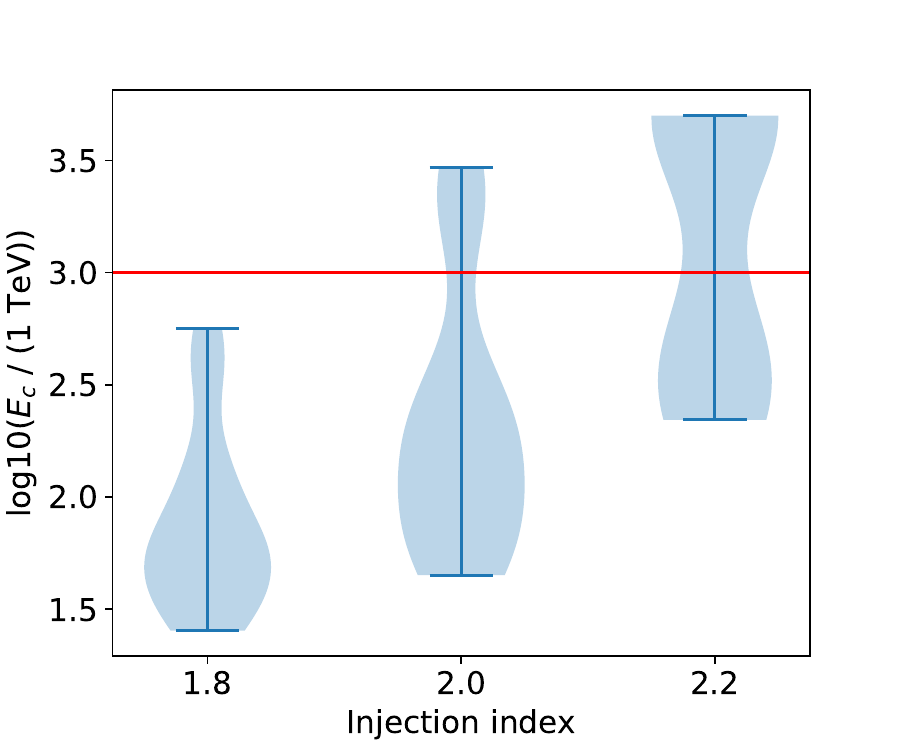}
\includegraphics[width=0.49\textwidth]{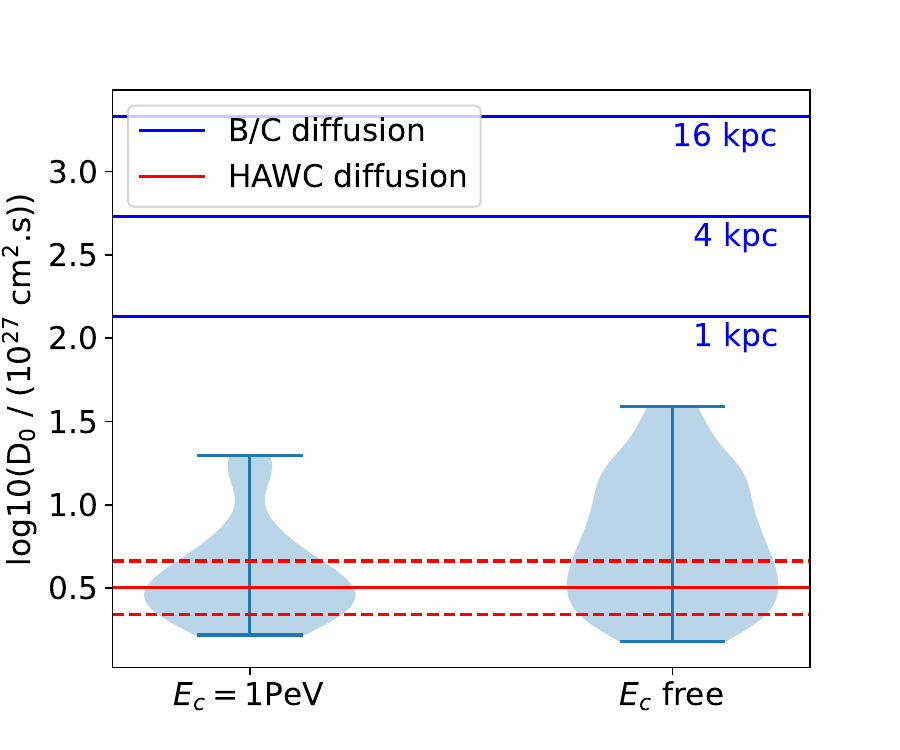}
\caption{Distribution of best-fit values for all parameter combinations resulting in a p-value $<0.003$. Left: correlation between injection index and cut-off energy when $E_c$ is left free. Right: Best-fit diffusion coefficient in the case of an energy cut-off fixed to 1\,PeV and left free to vary. }
\label{fig:pvalue_violin}
\end{figure}

Figure \ref{fig:pvalue_violin} shows the distribution of fitted $E_c$ for models with p-value $>0.003$; as expected this depends strongly on the assumed index $\alpha$ of the electron injection spectrum.  The best-fit normalisation for the diffusion coefficient is found to be systematically less than the Galactic value derived from the cosmic ray B/C ratio (Figure \ref{fig:pvalue_violin}). For both cases with $E_c = 1$\,PeV and $E_c$ as a free parameter of the fit, the majority of models favour a $D_0$ value (at an electron energy of 100\,TeV) consistent with that obtained in \cite{hawc}. %We obtain a value of $D_0 = 7.6^{+1.5}_{-1.2} \times 10^{27}$ cm$^2$s$^{-1}$ 

\begin{figure}
\centering
\includegraphics[width=0.49\textwidth]{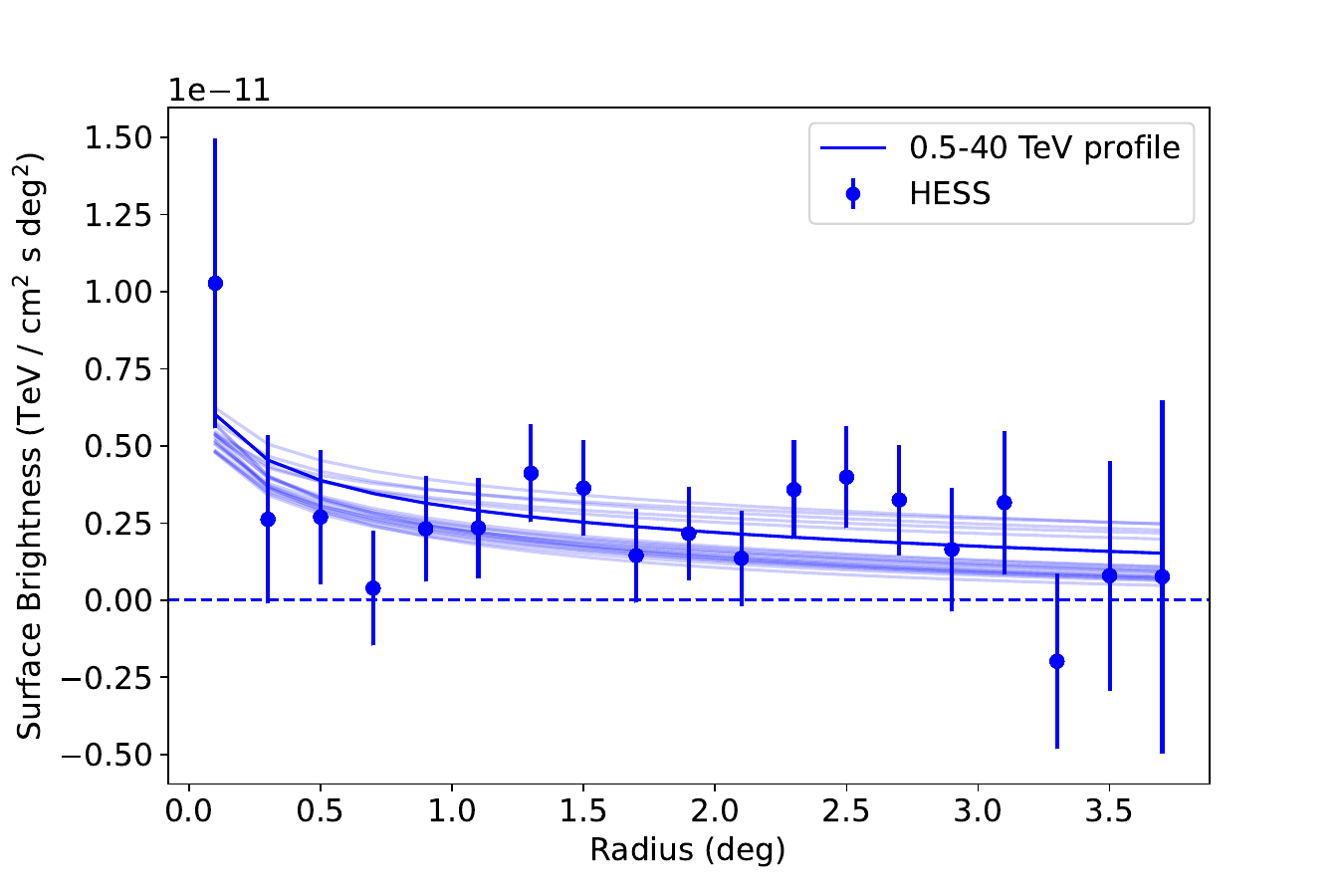}
\includegraphics[width=0.49\textwidth]{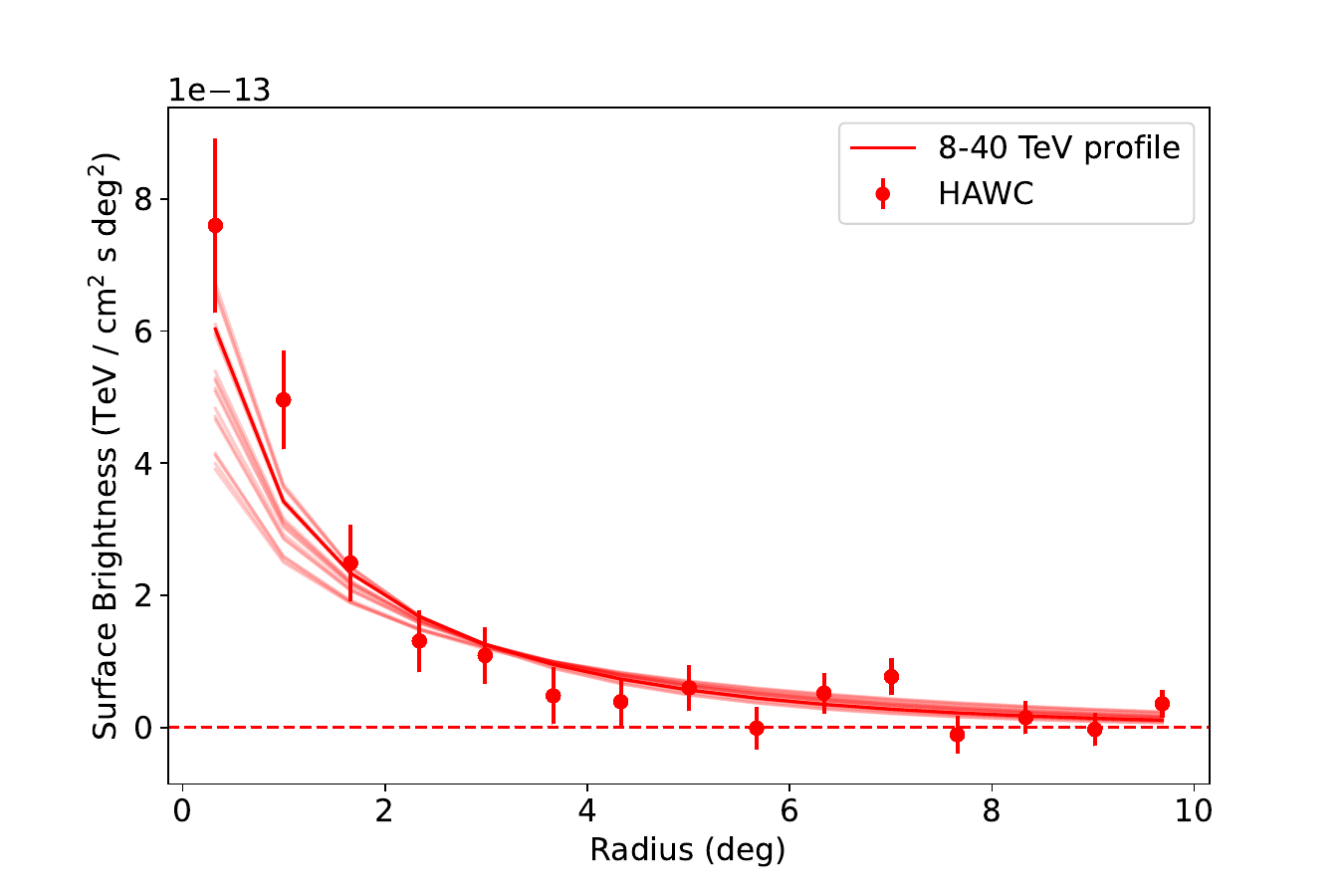}
\includegraphics[width=0.85\textwidth]{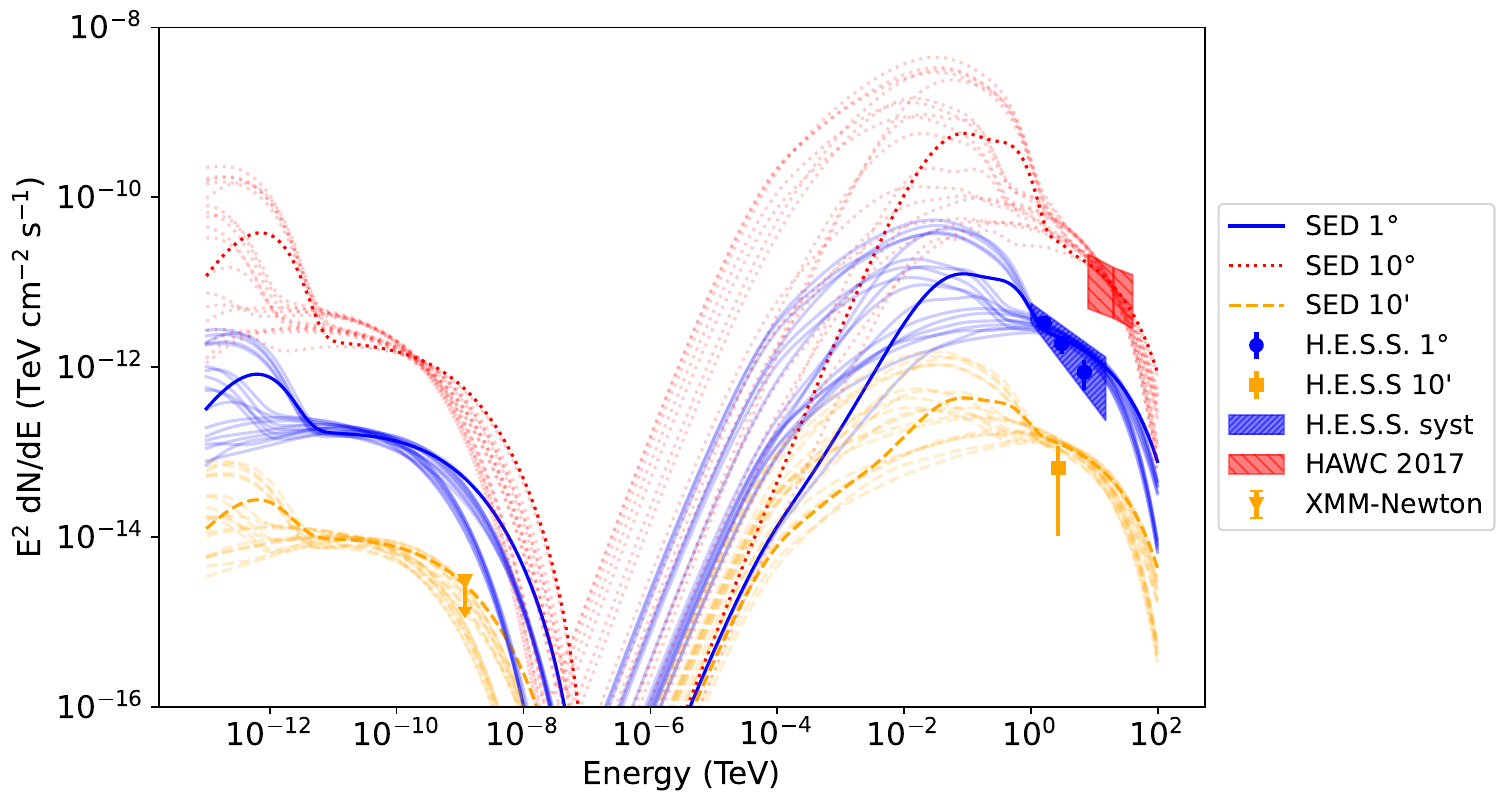}
\caption{Diffusion model jointly fit to the HAWC, H.E.S.S. and XMM-Newton data. For the two ground-based instruments radial profiles on degree scales are provided. The XMM-Newton upper limit is extracted from a 10' radius region, around the pulsar, from which a corresponding H.E.S.S. flux point is also extracted. }
\label{fig:modelfit}
\end{figure}

\begin{figure}
\centering
\includegraphics[width=0.49\textwidth]{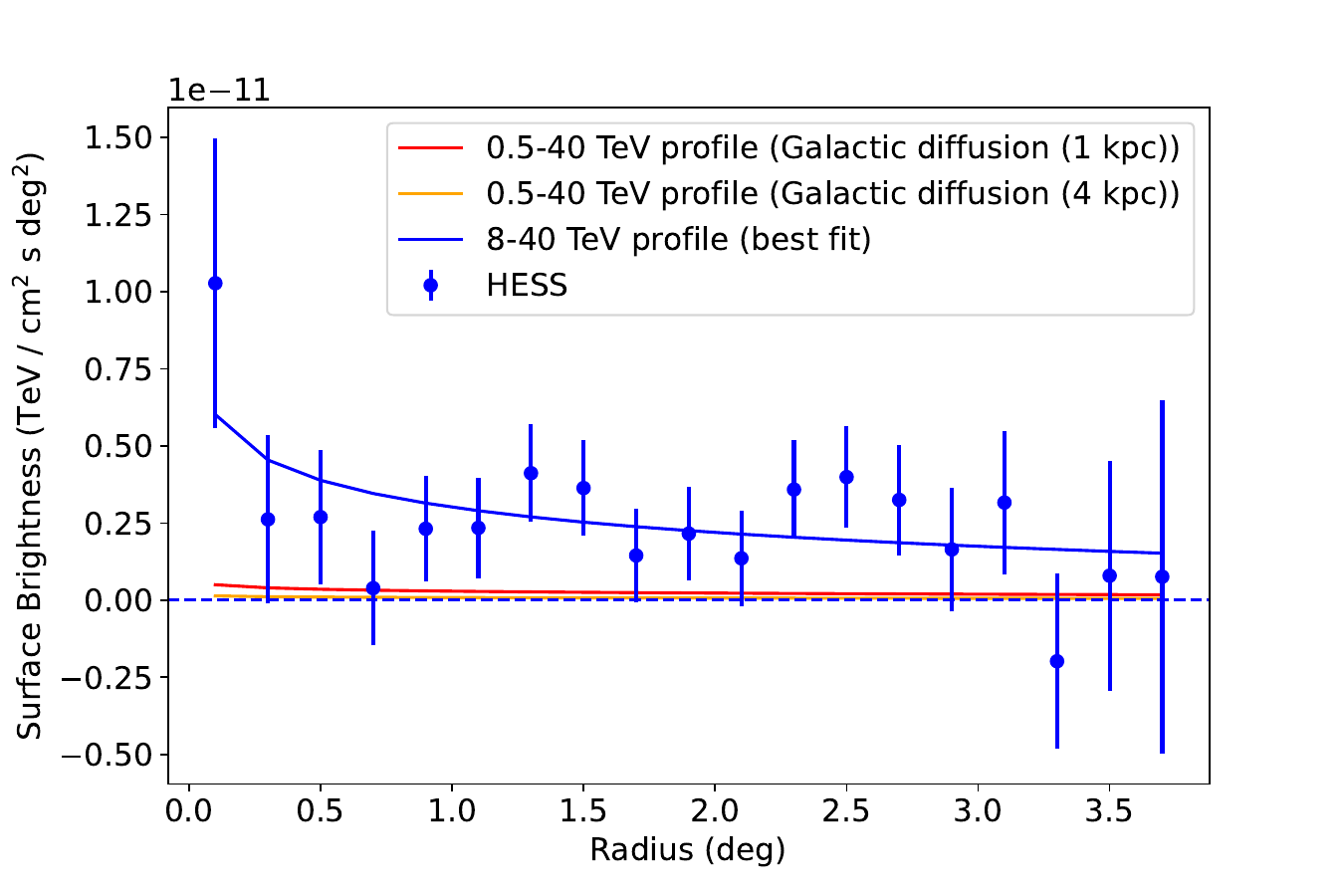}
\includegraphics[width=0.49\textwidth]{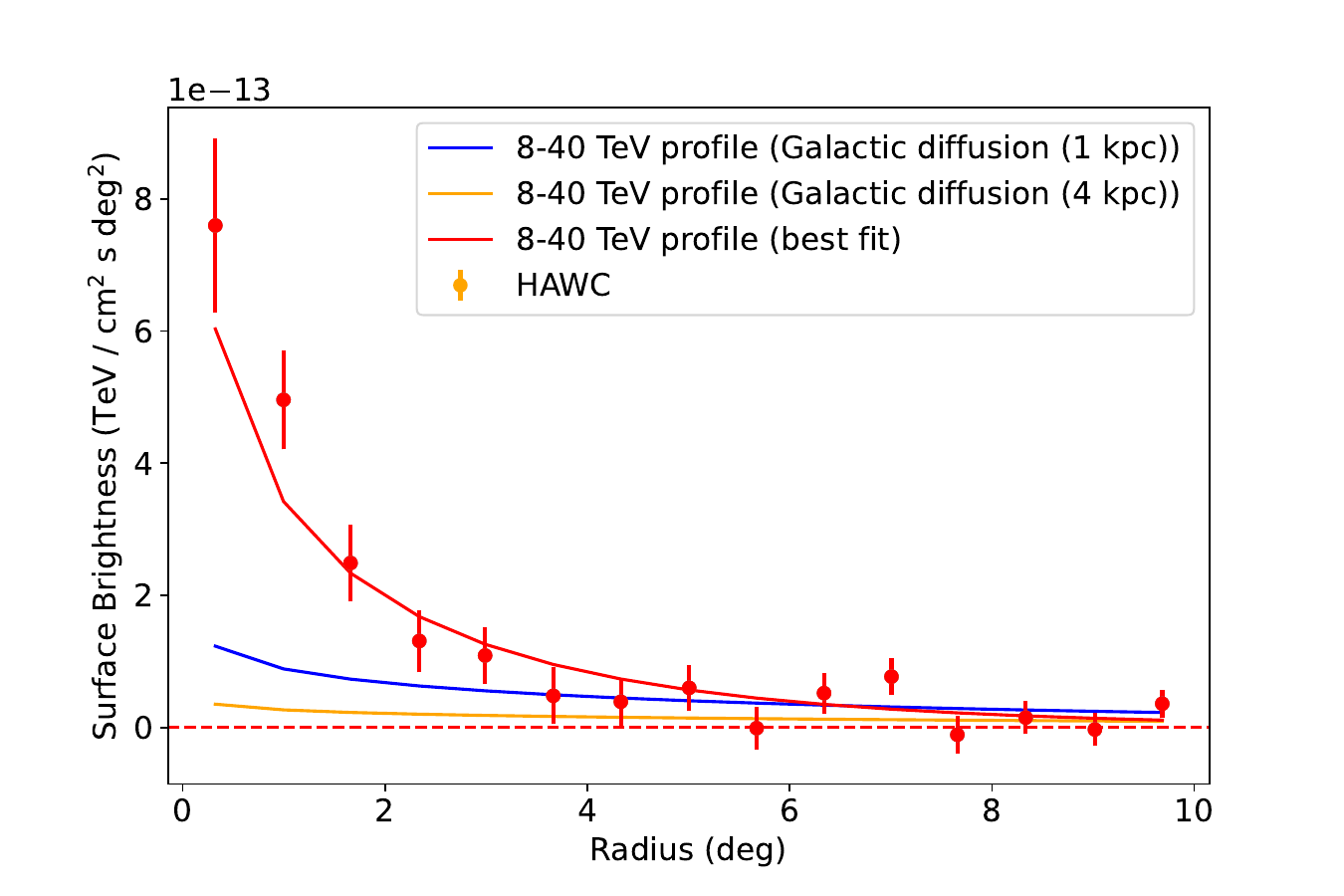}
\includegraphics[width=0.85\textwidth]{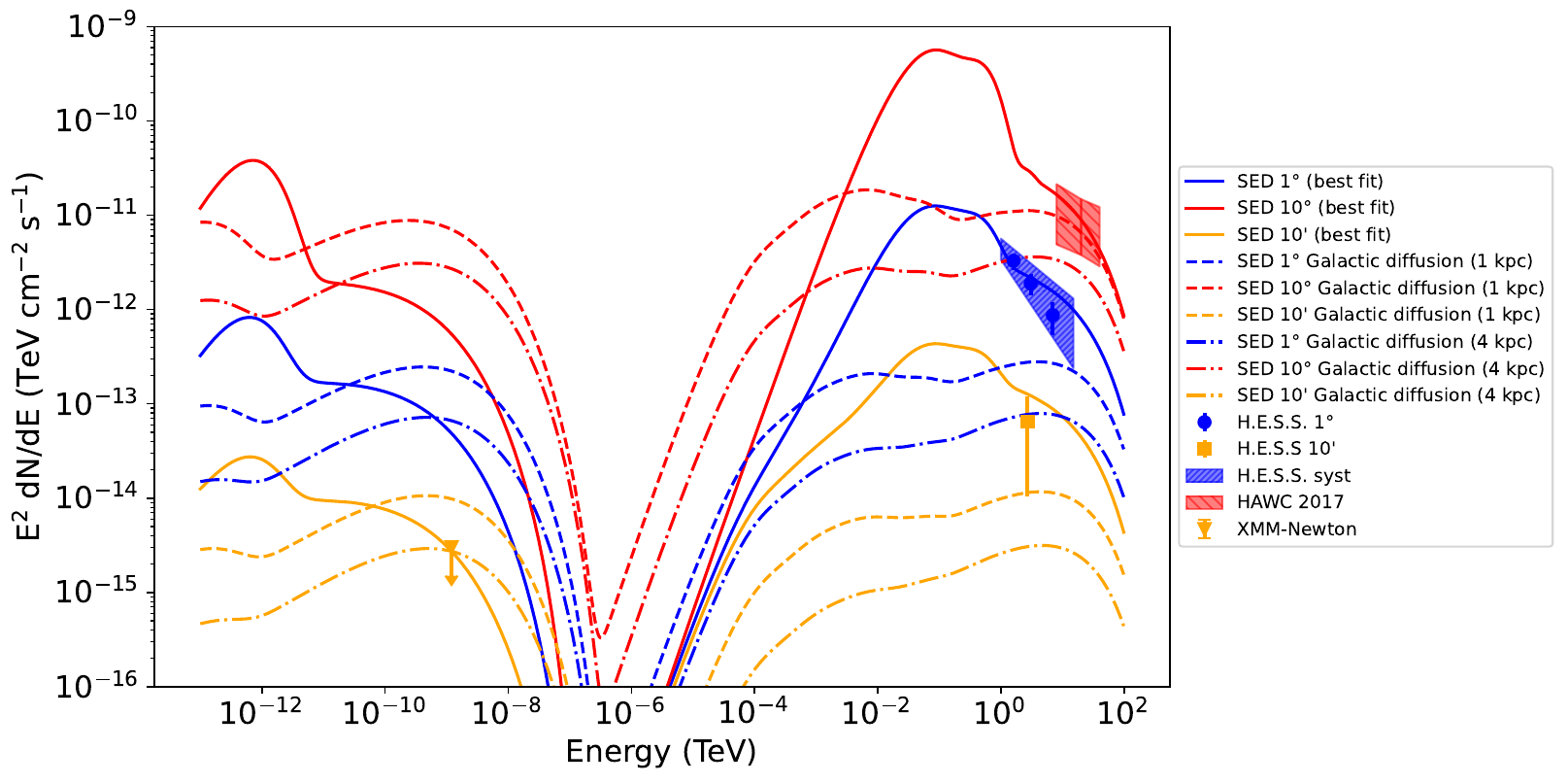}
\caption{Best fit model jointly fit to the HAWC, H.E.S.S. and XMM-Newton data. For the two ground-based instruments radial profiles on degree scales are provided. The XMM-Newton upper limit is extracted from a 10' radius region, around the pulsar, from which a corresponding H.E.S.S. flux point is also extracted. The SEDs that is obtained for typical galactic diffusion parameters are displayed as well, under the assumption of a diffusive halo height of $1$ and $4$ kpc.}
\label{fig:modelGalactic}
\end{figure}

Figure \ref{fig:modelfit} shows the model curves for all models with a p-value $>0.003$. The highlighted curve represents the model with the highest overall p-value of 0.37,  corresponding to the parameter values $n = 4.5$, $\eta = 0.1$, $\alpha = 1.8$, $\delta = 1.0$, $B = 1\mu$G and with fitted parameters $D_0 = 7.6^{+1.5}_{-1.2} \times 10^{27}$ cm$^2$s$^{-1}$ and $E_c = 74^{+17}_{-11}$ TeV.
Figure \ref{fig:modelGalactic} shows a comparison between the best fit and a model curve assuming a scenario in which the diffusion coefficient normalisation adopts a typical galactic average value. This galactic diffusion scenario is defined as $n = 3$, $\eta = 0.5$, $\alpha = 1.8$, $E_c = 74$ TeV, $\delta = 0.3$, $B = 3\,\mu$G \cite{Delahaye10} and $D_0$ fixed to B/C diffusion values obtained under different assumptions of the diffusive halo height \cite{genolini}.

\section{Conclusion}

With this work, we show that the $\gamma$-ray emission detected by H.E.S.S. in the vicinity of the Geminga pulsar \cite{hess} is consistent with that measured by \cite{hawc} in preferring a normalisation of the diffusion coefficient considerably below the galactic average. The detectability of extended $\gamma$-ray emission around the Geminga pulsar for both H.E.S.S. and HAWC would have been impossible in a case of a faster diffusion such as that expected in the galactic diffusion scenario. The discrepancy is particularly clear when the model with typical galactic diffusion values is directly compared with data. Based on our investigation of X-ray upper limits within a $10'$ region surrounding the pulsar, we can draw the conclusion that in a scenario involving a single diffusion zone, and assuming a constant magnetic field spanning the X-ray to $\gamma$-ray range, the magnetic field must be less than $1\,\mu G$  in the absence of a sub-PeV energy cut-off. To account for a magnetic field of $1\,\mu G$, a lower energy cut-off below $75$\,TeV is necessary.

In conclusion, a scenario comprising galactic-like diffusion and magnetic field properties in the vicinity of Geminga would imply that the halo of electrons would be undetectable in VHE $\gamma$-rays by both HAWC and H.E.S.S., and potentially detectable in X-ray. Observational evidence now indicates that the converse is actually the case, hence the modelling results are consistent with a diffusion coefficient considerably below galactic average values in the vicinity of the Geminga pulsar. 

%More stuff here. Expand references. 

%% Full authors list (ONLY FOR COLLABORATIONS)
\clearpage
\section*{Full Authors List: H.E.S.S. Collaboration}

\scriptsize
\noindent
F.~Aharonian$^{1,2,3}$, 
F.~Ait~Benkhali$^{4}$, 
A.~Alkan$^{5}$, 
J.~Aschersleben$^{6}$, 
H.~Ashkar$^{7}$, 
M.~Backes$^{8,9}$, 
A.~Baktash$^{10}$, 
V.~Barbosa~Martins$^{11}$, 
A.~Barnacka$^{12}$, 
J.~Barnard$^{13}$
% ...

\medskip

\noindent
$^{1}$Dublin Institute for Advanced Studies, 31 Fitzwilliam Place, Dublin 2, Ireland\\
$^{2}$Max-Planck-Institut f\"ur Kernphysik, P.O. Box 103980, D 69029 Heidelberg, Germany\\
$^{3}$Yerevan State University,  1 Alek Manukyan St, Yerevan 0025, Armenia\\
$^{4}$Landessternwarte, Universit\"at Heidelberg, K\"onigstuhl, D 69117 Heidelberg, Germany\\
$^{5}$IRFU, CEA, Universit\'e Paris-Saclay, F-91191 Gif-sur-Yvette, France\\
$^{6}$Kapteyn Astronomical Institute, University of Groningen, Landleven 12, 9747 AD Groningen, The Netherlands\\
$^{7}$Laboratoire Leprince-Ringuet, École Polytechnique, CNRS, Institut Polytechnique de Paris, F-91128 Palaiseau, France\\
$^{8}$University of Namibia, Department of Physics, Private Bag 13301, Windhoek 10005, Namibia\\
$^{9}$Centre for Space Research, North-West University, Potchefstroom 2520, South Africa\\
$^{10}$Universit\"at Hamburg, Institut f\"ur Experimentalphysik, Luruper Chaussee 149, D 22761 Hamburg, Germany\\
$^{11}$Deutsches Elektronen-Synchrotron DESY, Platanenallee 6, 15738 Zeuthen, Germany\\
$^{12}$Obserwatorium Astronomiczne, Uniwersytet Jagiello{\'n}ski, ul. Orla 171, 30-244 Krak{\'o}w, Poland\\
$^{13}$Department of Physics, University of the Free State,  PO Box 339, Bloemfontein 9300, South Africa\\
% ...

%\noindent \textbf{Note comment afterwards:} Collaborations have the possibility to provide an authors list in xml format which will be used while generating the DOI entries making the full authors list searchable in databases like Inspire HEP. \\
%
%\scriptsize
%\noindent
%first.author$^1$, 
%second.author$^2$, 
%third.author$^3$ % .... more names
%and 
%last.author$^{n}$ \\
%
%\noindent
%$^1$first.affiliation.
%$^2$second.affiliation. % .... more affiliation
%$^{m}$last.affiliation.


\begin{thebibliography}{99}
\bibitem{Milagro07} A. A. Abdo et. al. ApJ {\bf 700} (2009) L127
\bibitem{hawc} Abeysekara et. al. Science {\bf 358} (2017) p. 911-914 
%\vspace[-0.4cm]
\bibitem{Geminga_Xray} P. A. Caraveo et. al. Science {\bf 301} (2003) p. 1345-1347 
\bibitem{Giacinti_halo} G. Giacinti et. al. A\&A {\bf 636} (2020) A113
\bibitem{Linden2017PhRvD..96j3016L} T. Linden et. al. PRD {\bf 96} (2017) 103016
\bibitem{psrhaloreview} L\'opez-Coto et. al. Nature Astronomy {\bf 6} (2022) 199
\bibitem{Evoli} C. Evoli, T. Linden, G. Morlino, PRD {\bf 98} (2018) 063017 
\bibitem{hess} H.E.S.S. Collaboration, A\&A {\bf 673} (2023) A148
\bibitem{hawchess} Abdalla et. al., ApJ {\bf 917} (2021) 6
%\vspace[-0.4cm]
\bibitem{xmm} Liu et. al. ApJ {\bf 875} (2019) 149
\bibitem{fermi_diMauro} M. di\,Mauro, S. Manconi \& F. Donato, PRD {\bf 100} (2019) 123015
\bibitem{Delahaye10} T. Delahaye et. al. A\&A {524} (2010) A51
\bibitem{genolini} Y. Genolini et. al. PRD {\bf 99} (2019) 123028 
\end{thebibliography}
\end{document}